\documentclass[12pt]{iopart}
\newcommand{\lap}[0]{\int^{-\infty}_{0} dp \hspace*{1mm} e^{-pz} \hspace*{1mm}}
\newcommand{\sig}[1]{\sum^{\infty}_{#1=0}}

\newcommand{\sh}[1]{\sinh\sqrt{k}#1}
\newcommand{\ch}[1]{\cosh\sqrt{k}#1}
\usepackage{amssymb}
\usepackage[dvips]{graphicx}

\begin{document}

\title[Reconnection of Stable/Unstable Manifolds of the Harper Map]
{Reconnection of Stable/Unstable Manifolds of the Harper Map 
\\ $-$ {\large
Asymptotics-Beyond-All-Orders Approach} $-$}

\author{Shigeru Ajisaka\dag\ 
\footnote[3]{To
whom correspondence should be addressed (g00k0056@suou.waseda.jp)}
and Shuichi Tasaki\dag  
}

\address{\dag\ Advanced Institute for Complex Systems
and 
Department of Applied Physics,\\
School
of Science and Engineerings,
Waseda University,\\
3-4-1 Okubo, Shinjuku-ku,
Tokyo 169-8555,
Japan}

\begin{abstract}
The Harper map is one of the simplest chaotic systems exhibiting 
reconnection of invariant manifolds. 
The method of asymptotics beyond all orders (ABAO) is used to construct 
stable/unstable manifolds of the Harper map. By enlarging the neighborhood 
of a singularity, the perturbative solution of the unstable manifold is 
expressed as a Borel summable asymptotic expansion in a sector including 
$t=-\infty$ and is analytically continued to the other sector, where 
the solution acquires new terms describing heteroclinic tangles.
When the parameter changes to the reconnection threshold, the 
stable/unstable manifolds are shown to acquire new oscillatory portion 
corresponding to the heteroclinic tangle after the reconnection. 
\\
\\
Mathmatics Subject Classification:
34C37,\ 37C29,\ 37E99,\ 40G10,\ 70K44
\end{abstract}



\maketitle

\section{Introduction}

Bifurcation involving chaotic motions is one of the origins of 
diversity in complex systems and reconnection among stable/unstable 
manifolds is such an example.
One of the simplest systems exhibiting reconnection is the
Harper map~\cite{Saito,Shinohara}. 
For practical applications, it is helpful to have an analytical view
and we analytically study the reconnection in the nearly integrable 
Harper map.

To study weakly perturbed stable/unstable manifolds, the conventional 
Melnikov perturbation method is not applicable since splitting 
between stable and unstable manifolds is exponentially 
small for small perturbation parameter $\sigma$.
The method of asymptotics beyond all orders (ABAO method) is one of
the useful methods dealing with such situations. 
The key idea is to employ the so-called inner equation, which magnifies
the behavior of the solution near its singularities, and to
apply the Borel transformation for investigating the inner equation
and analytically continuing its solution. 
The idea of the inner equation was first used by Lazutkin, Schachmannski,
Tabanov\cite{Lazutkin} 
to derive the first crossing angle between the stable and unstable 
manifolds of Chirikov's standard map and by Kruskal and Segur\cite{Kruskal} 
to study a 
singular perturbation problem of ordinary differential equations. 
Hakim and Mallick\cite{Hakim} introduced the technique of Borel
transformation within in this context and obtained the first crossing
angle previously obtained in \cite{Lazutkinf}.
Tovbis, Tsuchiya and Jaff\'e~\cite{Tovbis} improved their method and
derived analytical approximations of perturbed stable/unstable 
manifolds for the H\'enon map. 
In these approaches, only the dominant part of the inner solution
is taken into account. 
Later, Nakamura and Kushibe~\cite{NakamuraKushibe} proposed a method
of systematically improving the solution of the inner equation with
the aid of the Stokes multiplier and studied Chirikov's standard map.
The ABAO method was applied to some higher dimensional systems
as well\cite{Hirata1,Hirata2}.
Also, Gelfreich and his collaborators\cite{Gelfreich1,Gelfreich2,Gelfreich3,Gelfreich4,Gelfreich5} are studying the 
splitting of separatrices and related bifurcations for various
systems by a slightly different but more rigorous way. 

Roughly speaking, the procedure\cite{Tovbis,NakamuraHamada,Nakamurabook,NakamuraKushibe} obtaining 
an approximate expression of
the unstable manifold is summarized as follows:
\begin{enumerate}

\item[(i)] Find singularities in the complex time domain of the lowest order 
ordinary perturbative solution of the unstable manifold.

\item[(ii)] Magnify the neighborhood of a singularity closest to the
real axis and derive an asymptotic expansion of the lowest order
solution which is valid in a sector containing $t=-\infty$.

\item[(iii)] The method of the Borel transformation is used to
construct the asymptotic expansion which is valid in the other sector.
Usually, there appear additional terms which are exponentially small 
for real time values.

\item[(iv)] Going back to the original equation, derive equations 
for corrections corresponding to the exponentially small terms found 
in the previous step. An appropriate solution is chosen by matching 
its asymptotic expansion with that obtained in the previous step.

\end{enumerate}
The additional terms are determined by the singularities of the 
Borel-transformed solutions and, so far, only contributions from
poles were considered.
In this paper, with the aid of ABAO method, the reconnection of the
unstable manifold is studied for the Harper map. Our analysis mainly 
follows the method by Nakamura and Kushibe~\cite{NakamuraKushibe},
but contributions from the branch cuts of the Borel-transformed
solution are also taken into account. 

\begin{figure}[b]
\label{fig:differential}
\begin{center}
\includegraphics[width=12cm,keepaspectratio,clip]{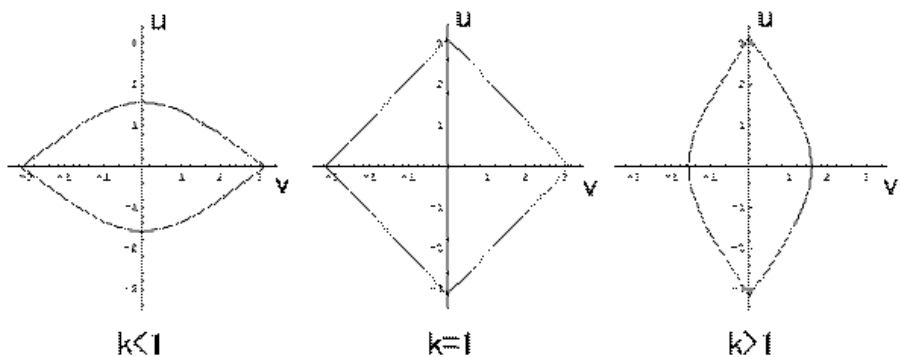}
\caption{Separatrix of the Harper map in the continuous limit}
\end{center}
\end{figure}

The Harper map depends on a real parameter $k$ and is defined 
on $(v,u)~\in~[-\pi,\pi]^2$:
\begin{eqnarray}
v(t+\sigma)-v(t)&=&-\sigma \sin u(t)
\nonumber \\
u(t+\sigma)-u(t)&=&k\sigma \sin v(t+\sigma)
\label{eq:Harper}
\end{eqnarray}
where $\sigma (>0)$ is the time step and plays a role of the small
parameter. 
Since the case of $k<0$ is conjugate to that of $k>0$, it is 
sufficient to consider the latter~\cite{Shinohara}.
In the continuous time limit, where $\sigma\to 0$, the map
reduces to a set of integrable differential equations:
\begin{eqnarray}
v'(t)&=&- \sin u(t) \nonumber
\\
u'(t)&=&k \sin v(t) \ . \label{eq:Harper0}
\label{eq:Harper0}
\end{eqnarray}
Throughout this paper, The prime is used to indicate the differentiation
with respect to time $t$ such as $u'(t)\equiv{\displaystyle du(t)\over
\displaystyle dt}$.
Eq.(\ref{eq:Harper0}) admits topologically different separatrices 
depending on the parameter $k$ (cf.~Fig.~\ref{fig:differential})
and the separatrix changes its shape when $k\to 1$.
Since the solution for $k>1$ can be obtained from that for $k<1$ by a simple 
symmetry argument (cf. \ref{appC}), we restrict ourselves to the case 
of $0<k<1$ and consider the change of the stable/unstable manifolds 
for $k\to 1$. In the text, an approximation of the unstable manifold
is constructed and only the result of the approximate stable manifold 
is given. 

The rest of this paper is arranged as follows: In the next section, 
the lowest order solution of the Melnikov perturbation method is 
constructed and the inner equation is derived. 
In Sec.~3, the Borel resummation method is used to solve the inner
equation and derive the additional exponentially small terms. 
In Sec.~4, the solution of the original equation with the corresponding 
corrections is obtained and compared with numerical calculations.
The reconnection of the unstable manifolds is discussed in Sec.~5.
Sec.~6 is devoted to the summary.

\section{Melnikov Perturbation and Inner Equation}
Let $v_{0},u_{0}$ be the solutions obtained by Melnikov perturbation, i.e.,
$$
v_{0}(t)\equiv\sig{n}\sigma^{n} v_{0n}(t)
\ ,\qquad 
u_{0}(t)\equiv\sig{n}\sigma^{n} u_{0n}(t)
$$
The lowest order terms satisfy (\ref{eq:Harper0})
and the first and second order terms obey 
\begin{eqnarray}
&&v'_{01}(t)+u_{01}(t)\cos u_{00}(t)=-\frac{1}{2}v''_{00}(t)
\nonumber
\\
&&u'_{01}(t)- kv_{01}(t)\cos v_{00}(t)=\frac{1}{2}u''_{00}(t) 
\label{Meq:1st}
\end{eqnarray}

\begin{eqnarray}
&&v'_{02}(t)+u_{02}(t)\cos u_{00}(t) 
= -\frac{1}{2}v''_{01}(t)-\frac{1}{6}v'''_{00}(t)
+\frac{u^{2}_{01}(t)}{2}\sin u_{00}(t)
\nonumber
\\
&&u'_{02}(t) -k v_{02}(t) \cos v_{00}(t)
= \frac{1}{2}u''_{01}(t)-\frac{1}{6}u'''_{00}(t)
-k \frac{v_{01}^{2}(t)}{2}
\sin v_{00}(t)
\label{Meq:2nd}
\end{eqnarray}

The lowest order solution of the 
unstable manifold with $v_{0}(0)=0$ is given by
\begin{eqnarray}
v_{00}(t)
&=&-2\tan^{-1}
\left[
\sqrt{1-k} \sinh \sqrt{k}t
\right]
\nonumber
\\
u_{00}(t)
&=&2\tan^{-1}
\left[
\sqrt{\frac{k}{1-k}}
\frac{1}{\cosh \sqrt{k}t}
\right]
\label{sol:0th}
\end{eqnarray}
Because of the boundary conditions $v_{0n}(t) \to 0, 
\ u_{0n}(t) \to 0$ for $t\to -\infty$
and $v_{0n}(0)=0$, the first and the second order solutions are
\begin{eqnarray}
v_{01}(t)=0
\ , \qquad
u_{01}(t)=\frac{1}{2}y_{1}(t)
\label{sol:1}
\end{eqnarray}
\begin{eqnarray}
v_{02}(t)&=&
-\frac{1}{24}\left[
x_{1}'(t)+x_{1}(t)\left(kt-2(1-k)
\frac{\sqrt{k}\tanh\sqrt{k}t}{1-k\tanh^{2}\sqrt{k}t}\right)
\right]
\nonumber
\\
u_{02}(t)&=&
\frac{1}{24}\left[
2y_{1}'(t)-y_{1}(t)\left(kt-2(1-k)
\frac{\sqrt{k}\tanh\sqrt{k}t}{1-k\tanh^{2}\sqrt{k}t}\right)
\right]
\end{eqnarray}
where auxiliary functions are given by
\begin{eqnarray}
x_{1}(t)&=&-2\sqrt{k(1-k)}\frac{\ch{t}}{1+(1-k)\sinh^{2}\sqrt{k}t}
\nonumber
\\
y_{1}(t)&=&-2k\sqrt{1-k}\frac{\sh{t}}{1+(1-k)\sinh^{2}\sqrt{k}t}
\label{eq:x1}
\end{eqnarray}

\begin{figure}[t]\label{fig:singular}
\begin{center}
\includegraphics[width=6cm,keepaspectratio,clip]{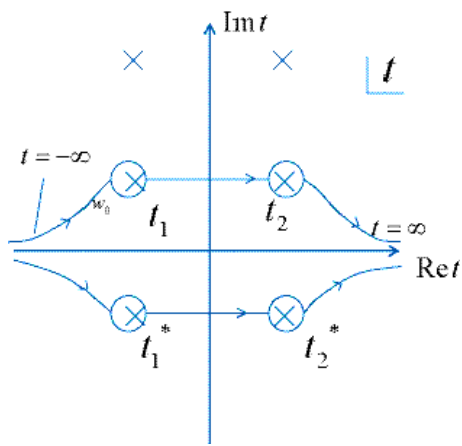}
\caption{Singular points (crosses) of the lowest order solution $v_{00}, \ u_{00}$ in the complex time domain.}
\end{center}
\end{figure}

We observe that the lowest order solution (\ref{sol:0th}) has two sequences of singular 
points in the complex $t$-plane (see Fig.~\ref{fig:singular}).
\begin{eqnarray}
t
=
\frac{1}{\sqrt{k}}\left\{
\ln\frac{1\pm\sqrt{k}}{\sqrt{1-k}}+\left(n+\frac{1}{2}\right)\pi i
\right\}
\end{eqnarray}

Now we examine the behavior of perturbative solutions near two singular points in the
upper half plane closest to the real axis:
$t_{1} \equiv {1\over \sqrt{k}}\left\{\ln\frac{1-\sqrt{k}}{\sqrt{1-k}}+\frac{1}{2}\pi i\right\}$,
$t_{2} \equiv {1\over \sqrt{k}}\left\{\ln\frac{1+\sqrt{k}}{\sqrt{1-k}}+\frac{1}{2}\pi i
\right\}$.
From now on, $t_{c}$ stands for $t_1$ or $t_2$, and the upper sign corresponds to 
$t_{1}$ and the lower one corresponds to $t_{2}$.

As easily seen, the perturbative solution admits an expansion like
\begin{eqnarray}
v_{0}(t,\sigma)-v_{00}(t)&=&\sum_{n=1}^{\infty}
\sig{l}\frac{a^{(n)}_{l+1}}{(t-t_{c})^{n-l}}\sigma^{n}
\label{div.exp}
\end{eqnarray}
This indicates that the higher order terms with respect to $\sigma$ has 
higher order poles at $t=t_{c}$ and that the behavior near the singularities
should be carefully investigated. For this purpose, it is convenient to introduce 
a rescaled 
variable $z \equiv \frac{t-{t_c}}{\sigma}$~\cite{Hakim,Tovbis,NakamuraHamada,Nakamurabook,NakamuraKushibe}.
Then, one finds that $\Phi_{0} (z,\sigma) \equiv v_{0}(t_c+\sigma z,\sigma)-v_{00}(t_c+\sigma z)$
and $\Psi_{0} (z,\sigma) \equiv u_{0}(t_c+\sigma z,\sigma)-u_{00}(t_c+\sigma z)$ can be 
expanded into power series with respect to $\sigma$:
\begin{eqnarray} 
\Phi_{0} (z,\sigma)=\sig{l}\Phi_{0l}(z) \sigma^{l} \ ,
\qquad
\Psi_{0} (z,\sigma)=\sig{l}\Psi_{0l}(z) \sigma^{l} \ ,
\label{sol:inner}
\end{eqnarray}
where $(\Phi_{00}, \Psi_{00})$, $(\Phi_{01}, \Psi_{01})$ $\dots$ correspond to the
most divergent terms, the next divergent terms $\dots$.

We remark that, in terms of 
$\Phi(z,\sigma)\equiv v(t_c+\sigma z,\sigma)-v_{00}(t_c+\sigma z)$ and 
$\Psi(z,\sigma)\equiv u(t_c+\sigma z,\sigma)-u_{00}(t_c+\sigma z)$,
the Harper map (\ref{eq:Harper}) reads as
\begin{eqnarray}
\Delta\Phi (z,\sigma)
&=& -\sigma
\sin\Psi(z,\sigma)\cos u_{00}(t_{c}+\sigma z)   \nonumber \\
&&~-\sigma\cos\Psi(z,\sigma)\sin u_{00}(t_{c}+\sigma z)
-\Delta v_{00}(t_{c}+\sigma z)
\nonumber
\\
\Delta\Psi (z-1,\sigma)
&=& k\sigma
\sin\Phi(z,\sigma)\cos v_{00}(t_{c}+\sigma z)  \label{eq:inner}
\\
&&+k\sigma\cos\Phi(z,\sigma)
\sin v_{00}(t_{c}+\sigma z)
-\Delta u_{00}(t_{c}+\sigma (z-1))
\nonumber
\end{eqnarray}
where $\Delta$ stands for the difference operator: $\Delta f(z)=f(z+1)-f(z)$,
and that 
 $\Phi_{0}(z,\sigma),\ \Psi_{0}(z,\sigma)$ are its
perturbative solutions with respect to $\sigma$.
This equation will be referred to as an inner equation.

Substituting (\ref{sol:inner}) into (\ref{eq:inner}), 
one can easily obtain the equations for $\Phi_{0l},\ \Psi_{0l}$.
The first two sets are as follows
\begin{eqnarray}
\Delta\Phi_{00}(z)&=&i\frac{e^{\mp i\Psi_{00}(z)}}{z}-i\ln\left(1+\frac{1}{z}\right)
\nonumber
\\
\Delta\Psi_{00}(z)&=&\mp i\frac{e^{i\Phi_{00}(z+1)}}{z+1}\pm i\ln\left(1+\frac{1}{z}\right)
\label{eq:inner0}
\\
\Delta\Phi_{01}(z)
&=&\pm\left[i\frac{k-1}{2}+\frac{\Psi_{01}(z)}{z}\right] e^{\mp i\Psi_{00}(z)}
\mp\frac{i}{2}(k-1)
\nonumber
\\
\Delta\Psi_{01}(z)&=&
\left[\pm\frac{\Phi_{01}(z+1)}{z+1}-i\frac{1-k}{2}\right]e^{i\Phi_{00}(z+1)}
+\frac{i}{2}(1-k)
\label{eq:inner1}
\end{eqnarray}

By matching the residues at $z=0$:
\begin{eqnarray}
{\rm Res} \left.\pmatrix{\Phi_{0k}(z) \cr \Psi_{0k}(z)}\right|_{z=0}
=
{\rm Res} \left.\pmatrix{v_{0\ \mskip -5mu k+1}(t) \cr u_{0\ \mskip -5mu k+1}(t)}\right|_{t=t_{c}}
\ ,
\end{eqnarray}
one obtains the solution corresponding to the expansion (\ref{div.exp})
\begin{eqnarray}
\Phi_{00}(z)&=&\ \ \ \ \frac{i}{12z^{2}}-\frac{107i}{4320z^{4}}+{\rm O}\left(\frac{1}{z^5}\right)
\nonumber
\\
\Psi_{00}(z)&=&\mp\left(
\frac{i}{2z}-\frac{i}{24z^{2}}-\frac{i}{24z^{3}}+\frac{191i}{8640z^{4}}
\right)+{\rm O}\left(\frac{1}{z^5}\right)
\label{sol:inner 00}
\\
\Phi_{01}(z)&=&\ \ \ \ \mp\frac{i}{24}\frac{\pm kt_{c}+1}{z}+{\rm O}\left(\frac{1}{z^3}\right)\nonumber
\\
\Psi_{01}(z)&=&\frac{i(k-1)}{4}+\frac{i}{24}\frac{k(\pm t_{c}+1)}{z}-\frac{i}{48}
\frac{k(\pm t_{c}+1)}{z^{2}}+{\rm O}\left(\frac{1}{z^3}\right)
\label{sol:inner 01}
\end{eqnarray}

This is the asymptotic solution of the inner equation which is valid in the sector
involving Re$~z =-\infty$. It is necessary to analytically continue it to another
sector involving Re$~z =+\infty$ in order to construct an approximation of the
unstable manifold in the time domain Re$~t >$ Re$~t_c$. 
This will be carried out in the next section.

\section{Borel Transformation and Solution of Inner Equation}

\subsection{Borel Transformation and Analytic Continuation}
As mentioned in the previous section, in order to construct an approximation of the
unstable manifold in the time domain Re$~t >$ Re$~t_c$, the solution of the inner
equation in the sector involving Re[$z]=-\infty$ should be analytically continued 
to the sector involving Re[$z]=+\infty$. The analytic continuation is carried out
with the aid of the Borel transformation. Let us begin with $\Phi_{00}(z),\Psi_{00}(z)$.

\begin{figure}[t]
\begin{center}
\includegraphics[width=4cm,keepaspectratio,clip]{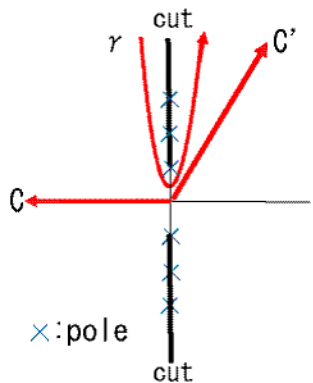}
\caption{Integral path in $p$-plane}
\label{fig:int}
\end{center}
\end{figure}

Since we are interested in the unstable manifold, we define the Borel transforms $V_0(p),\ U_0(p)$ 
of $\Phi_{00}(z),\Psi_{00}(z)$, respectively, as
\begin{eqnarray}
\Phi_{00}(z)\equiv L[V_0(p)](z) \equiv \lap V_0(p)\nonumber
\\
\Psi_{00}(z) \equiv L[U_0(p)](z) \equiv \lap U_0(p)
\label{borel trans}
\end{eqnarray}
The analytic continuations of $\Phi_{00}(z)$ and $\Psi_{00}(z)$ from Re$[z]<0$, 
Im$[z]<0$ to Re$[z]>0$ are obtained by rotating the integral path in the $p$-plane from C to C$'$
(cf. Fig.~\ref{fig:int}). Other possible singularities and branch cuts of $V_{0},\ U_{0}$ 
are also shown in Fig.~\ref{fig:int}.
Note that the integral path should be rotated over $\frac{\pi}{2}$~[rad] in a 
clockwise way so that the convergence domain rotates counterclockwise.
Thus, the analytic continuation from Re$[z]<0$, Im$[z]<0$ to $\pi+\epsilon<{\rm arg}[z]<2\pi+\epsilon$ is given by
\begin{eqnarray}
\label{analytic}
\int^{-\infty}_{0} dp \hspace*{1mm} e^{-pz} \hspace*{1mm}
V_0(p)
\to
\int^{-\infty e^{i(\pi/2+\epsilon)}}_{0} dp \hspace*{1mm} e^{-pz} \hspace*{1mm}
V_0(p)
-\int_{\gamma} dp \hspace*{1mm} e^{-pz} \hspace*{1mm}
V_0(p)
\end{eqnarray}
where $\epsilon$ is a small positive real number.

If one can write down the Borel-transformed inner solution explicitly, 
one has only to change the integral path.
But it is difficult and we have to deal with the inner equation and the 
Borel-transformed inner equation simultaneously.
The procedure can be summarized as follows

\begin{enumerate}

\item[(a)] 
Write down the equation for $V_0, \ U_0$:
\begin{eqnarray}
-i(e^{-p}-1) V_{0}(p) &=&1+ \int_{0}^{p} dp \left\{\sum_{n=1}^{\infty}
\frac{(\mp i)^{n}U_{0}^{(*n)}(p)}{n!}
-\frac{1-e^{-p}}{p} \right\}
\nonumber
\\
\pm i(1-e^{p}) U_{0}(p) &=&1+ \int_{0}^{p}dp \left\{ \sum_{n=1}^{\infty}
\frac{i^{n}V_{0}^{(*n)}(p)}{n!}-\frac{e^{p}-1}{p} \right\}
\label{innerborel}
\end{eqnarray}
where $V_0^{(*n)}$ denote the $n$th convolution defined by
\begin{eqnarray}
V_0^{(*n)}(p)=\int^{p}_{0} dx V_{0}(p-x) V_0^{(*(n-1))}(x)
\end{eqnarray}
This equation indicates that the Borel transforms $V_{0}(p),\ U_{0}(p)$ may have 
singularities at $p=\pm 2\pi i$ and branch cuts starting from them (cf. Fig.~\ref{fig:int}).
This observation and the nonlinearity of the inner equation imply that additional terms 
like $-2\pi i$Res$\left.\left(V_0(p)e^{-pz}\right)\right|_{p=2\pi in}$ and those from
branch cuts may appear in Re$[z]>0$.

\item[(b)]
Observation (a) suggests the expansions
\begin{eqnarray}
\left.\Phi(z,\sigma)\right|_{\sigma=0}=
\sig{n}\Phi_{n0}(z)e^{-2\pi i nz}
\nonumber
\\
\left.\Psi(z,\sigma)\right|_{\sigma=0}=
\sig{n}\Phi_{n0}(z)e^{-2\pi i nz}
\label{new expansion}
\end{eqnarray}
where $\Phi_{n0},\ \Psi_{n0}$ consist of finite polynomials of $z$
and power series of $1/z$. 
Solve the equations for $\Phi_{n0},\ \Psi_{n0}$, which can be obtained
by substituting (\ref{new expansion}) to (\ref{eq:inner}) and 
comparing the same order terms with respect to $e^{-2\pi iz}$.
Note that $\Phi_{00}, \Psi_{00}$ are the solutions of (\ref{sol:inner 00}).

\item[(c)] With the aid of the formula
\begin{eqnarray}
-2\pi i(-z)^{j-1}e^{-2\pi inz}
&=&
-\int_{\gamma} dp \hspace*{1mm} e^{-pz} \hspace*{1mm}
\left[\frac{(j-1)!}{(p-2\pi ni)^{j}}\right]
\nonumber
\\
\frac{-2\pi i}{z^{j+1}}e^{-2\pi inz}&=&
-\int_{\gamma} dp \hspace*{1mm} e^{-pz} \hspace*{1mm}
\left[\frac{(p-2\pi in)^{j}\ln(p-2\pi in)}{j!}\right]
\nonumber
\end{eqnarray}
find $V_{0}(p),\ U_{0}(p)$ in the upper half $p$-plane from the following equation
$$
-\int_{\gamma} dp \hspace*{1mm} e^{-pz} \hspace*{1mm}
V_{0}(p)
=\sum_{n=1}^{\infty}\Phi_{n0}(z) e^{-2\pi inz} \ , \quad
-\int_{\gamma} dp \hspace*{1mm} e^{-pz} \hspace*{1mm}
U_{0}(p)
=\sum_{n=1}^{\infty}\Psi_{n0}(z) e^{-2\pi inz}
$$
In the above, we choose the branch cut of the logarithm along the positive imaginary axis as shown
in Fig.~\ref{fig:int}. Note that this choice fixes the Stokes line on the negative imaginary axis in the 
$z$-plane, which corresponds to the line in the original $t$-plane joining $t_{c}$ and its complex conjugate.

\item[(d)]
By the procedures explained so far, only the singularities in the upper half $p$-plane are taken into account.
The contributions from the singularities in the lower half plane are determined 
from the fact that $V_{0},\ U_{0}$ is pure imaginary when $z$ is real.

\item[(e)]
Determine $V_{0},\ U_{0}$ by comparing the MacLaurin expansion of the analytical guess obtained from
(a)-(d) and the numerical power series solution of the Borel transformed inner equation
(\ref{innerborel}). 

\end{enumerate}

\subsection{Determination of $V_{0},\ U_{0}$ }
In this subsection, following the steps (b)-(e) explained before, $V_{0},\ U_{0}$  are calculated.

The equations of $\Phi_{10},\Psi_{10}$ are given by
\begin{eqnarray}
\Delta\Phi_{10}(z)&=&\pm\Psi_{10}(z)\frac{e^{\mp i\Psi_{00}(z)}}{z}
\ , \qquad
\Delta\Psi_{10}(z)=\pm \Phi_{10}(z+1)\frac{e^{i\Phi_{00}(z+1)}}{z+1}
\label{inner10}
\end{eqnarray}
which have two independent solutions:
\begin{eqnarray}
\Phi^{(A)}_{10}&=& -\frac{1}{z}+\frac{23}{48z^{3}} +\cdots 
\ , \qquad
\Psi^{(A)}_{10}=\pm
\left(\frac{1}{z}-\frac{1}{2z^{2}}-\frac{1}{48z^{3}}\right)+\cdots 
\nonumber
\\
\Phi^{(B)}_{10}&=&z+\frac{1}{12z}-\frac{17}{1080z^{3}}+\cdots 
\ , \quad
\Psi^{(B)}_{10}=\pm
\left(z+\frac{1}{2}-\frac{1}{540z^{3}}\right)+\cdots 
\nonumber
\end{eqnarray}
and $\Phi_{10},\ \Psi_{10}$ are their linear combinations.
\begin{eqnarray}
\left[
\begin{array}{l}
\Phi_{10}
\\
\Psi_{10} 
\end{array}
\right]
=
c_{A}
\left[
\begin{array}{l}
\Phi^{(A)}_{10}
\\ 
\Psi^{(A)}_{10}
\end{array}
\right]
+
c_{B}
\left[
\begin{array}{l}
\Phi^{(B)}_{10}
\\ 
\Psi^{(B)}_{10}
\end{array}
\right]
\label{sol:inner10}
\end{eqnarray}

From (\ref{sol:inner10}), $V_{0}(p)$ is given by
\begin{eqnarray}
-\int_{\gamma} dp \hspace*{1mm} e^{-pz} \hspace*{1mm}
V_{0}(p)
&=&\sum_{n=1}^{\infty}\Phi_{n0} e^{-2\pi inz}
\nonumber
\\
&=& \Phi_{10} e^{-2\pi iz} + \Phi_{20} e^{-4\pi iz}+\cdots
\nonumber
\\
&\approx& \left[
c_{B}\left(z+\frac{1}{12z}\right)-\frac{c_{A}}{z}
+{\rm O}\left(\frac{1}{z^3}\right)
\right]e^{-2\pi iz}
\label{match}
\end{eqnarray}
and $U_0$ can be represented in a similar way.
As easily seen from (\ref{innerborel}), the coefficients $c_{A}$ and $c_{B}$ are imaginary number.

In order to determine the Borel transforms $V_0,\ U_0$ from the expansion (\ref{match}) 
by taking into account the singularities in the lower half plane, we introduce
auxiliary functions 
$f^{(R)}_{lj}(p),\ f^{(I)}_{lj}(p)$ satisfying the following two conditions:

\begin{itemize}
\item
$\displaystyle \mskip 40 mu
-\int_{\gamma} dp \hspace*{1mm} e^{-pz} \hspace*{1mm}
f^{(R)}_{lj}(p)=
z^{j}e^{-2\pi l iz}
\nonumber
\\
~\mskip 40 mu
-\int_{\gamma} dp \hspace*{1mm} e^{-pz} \hspace*{1mm}
f^{(I)}_{lj}(p)=
iz^{j}e^{
-2\pi l iz}
\nonumber
$

\item
$f^{(R)}_{lj}(p),\ f^{(I)}_{lj}(p)$ 
is pure imaginary when $p$ is real.
\end{itemize}

Especially, for non-negative $j$, $f^{(R)}_{1j}(p)\ f^{(I)}_{1j}(p)$ are given by

\begin{eqnarray}
f^{(R)}_{1j}(p)&=&\frac{j!}{2\pi i}
\left(\frac{1}{(p-2\pi i)^{j+1}}+\frac{(-1)^{j+1}}{(p+2\pi i)^{j+1}}\right)
\\
f^{(I)}_{1j}(p)&=&i\frac{j!}{2\pi i}
\left(\frac{1}{(p-2\pi i)^{j+1}}-\frac{(-1)^{j+1}}{(p+2\pi i)^{j+1}}\right)
\end{eqnarray}
And $f^{(I)}_{1,-1}$ is
\begin{eqnarray}
f^{(I)}_{1,-1}(p)
&=&
-\frac{1}{2\pi}
\left[\ln(p-2\pi i) - \ln(p+2\pi i)\right]
\end{eqnarray}

In terms of these functions $V_0$ is given by
\begin{eqnarray}
V_0(p)&=&\frac{c_B}{i} f^{(I)}_{11}(p)+\frac{1}{i}(\frac{c_B}{12}-c_{A})f^{(I)}_{1,-1}(p)
\nonumber
\\
&=& M +{1\over 4\pi^3}
\sig{n} \left(c_B(2n+2)
-{4\pi^2 (\frac{c_B}{12}-c_{A})\over 2n+1}\right)
(-1)^n \left({p\over 2\pi}\right)^{2n+1} 
\label{compare2V}
\end{eqnarray}
where the constant $M$ depends on the choice of the Riemann surface of the 
logarithm.
In a similar way, one finds 
\begin{eqnarray}
\pm U_0(p)&=&\frac{1}{i}\left(
c_B f^{(I)}_{11}(p)+c_{A}f^{(I)}_{1,-1}(p)+\frac{c_B}{2}f^{(I)}_{10}(p)
\right)
\nonumber \\
&=&  M' +{1\over 4\pi^3}
\sig{n} \left(c_B(2n+2)
-{4\pi^2 c_{A}\over 2n+1}\right)
(-1)^n \left({p\over 2\pi}\right)^{2n+1} 
\nonumber
\\
&&-\frac{c_B}{4\pi^2}\sig{n} (-1)^n \left({p\over 2\pi}\right)^{2n} 
\label{compare2U}
\end{eqnarray}
where $M'$ depends on the choice of the Riemann surface.

In order to check the validity of analytical guess (\ref{compare2V}) and (\ref{compare2U}) 
about singularities and to 
evaluate $c_A,\ c_B$, we derive power series solutions of (\ref{innerborel}) 
numerically and compare them with the power series expansions of the R.H.S. 
of (\ref{compare2V}) and (\ref{compare2U}).
By substituting the power series expansions:
\begin{eqnarray}
V_{0}(p)&\equiv&\sig{n}a_{n}p^{n}
, \quad
U_{0}(p)\equiv\sig{n}b_{n}p^{n}
; \quad a_{0}=0 , \quad b_{0}=\mp i\frac{1}{2}
\nonumber
\end{eqnarray}
into (\ref{innerborel}) and comparing term by term, one can determine 
the coefficients $a_{n},\ b_{n}$ recursively.
They have the following asymptotic forms.
\begin{eqnarray}
-ia_{2n+1}(-1)^{n}(2\pi)^{2n+1} &\rightarrow& 0.27893(2n+2)-\frac{0.417}{(2n+1)},\ \  {\rm as} \ \ n\rightarrow\infty
\nonumber
\\
&\equiv& A_{1}(2n+2)-\frac{A_{2}}{(2n+1)}
\nonumber
\\
\pm ib_{2n+1}(-1)^{n}(2\pi)^{2n+1} &\rightarrow& -0.27893(2n+2)+\frac{0.417}{(2n+1)},\ \  {\rm as} \ \ n\rightarrow\infty
\nonumber
\\
-ia_{2n}&=&0,\ \ \ \ \forall n
\nonumber
\\
\pm ib_{2n}(-1)^{n}(2\pi)^{2n} &\rightarrow& 0.87628 \equiv A_3
\label{numerical}
\end{eqnarray}
The numerical evaluations of $A_1,\ A_3$ are quite robust.
On the other hand, $A_2$ is sensitive to an error of $A_1$ 
since the coefficients of $1/(2n+1)$ are fitted after subtracting 
the leading terms.
However, we observe that the coefficients of $1/(2n+1)$ in 
$a_{2n+1}(-1)^{n}(2\pi)^{2n+1}$ and
$b_{2n+1}(-1)^{n}(2\pi)^{2n+1}$ 
are always the same. 
This observation suggests that the coefficients of $1/z$ in $\Phi_{10},\ \Psi_{10}$ are identical,
which implies
\begin{eqnarray}
A_{2}={c_A \over i\pi}={c_B\over 24i\pi}= \frac{A_{1}\pi^{2}}{6}
\end{eqnarray}
The present numerical estimation gives a consistent value of $0.417$ 
with this relation. 

According to (\ref{compare2V}), (\ref{compare2U}) and (\ref{numerical}), the asymptotic 
expansions of $a_{2n+1}(-1)^{n}(2\pi)^{2n+1}$ and
$b_{2n+1}(-1)^{n}(2\pi)^{2n+1}$ 
with respect to $n$ should have the same leading terms.
This is indeed the case.
Moreover, $A_3/A_1$ should be equal to $\pi$
and we have an excellent agreement:
$\displaystyle
A_3/A_1=3.14158\cdots
$.
Therefore, we can conclude that our guess about 
the poles and branch cuts in the $p$-domain is valid.
Then the constants $c_A, \ c_B$ are given by $c_A=i \pi A_2$ 
and $c_{B}=i 4\pi^{3}A_{1}$.

So far, we have considered the first three dominant terms in
$V_{0},\ U_{0}$. From the asymptotic expansions of (\ref{match}),
one finds that $V_{0},\ U_{0}$ involve $f^{(I)}_{1j}(p)$
\break
$(j \le -2)$
and $f^{(I)}_{lj}(p)$~$(l \ge 2)$. 
As easily seen, one has the following rough estimations:
\begin{eqnarray}
f_{lj}^{(\lambda)}(p) \sim \sig{n}\frac{1}{n^2(2\pi l)^{n}}p^n
\ \ (j \le -2) \ , \quad 
f_{lj}^{(\lambda)}(p) \sim \sig{n}\frac{n^{j}}{(2\pi l)^{n}}p^n
\ \ (j \ge -1) \nonumber
\end{eqnarray}
Therefore, if $[f_{lj}^{(\lambda)}]_n$ denotes the absolute value
of the coefficient of $p^n$, one has the following relation for 
large $n$:
\begin{eqnarray} 
[f_{11}^{(\lambda)}]_n 
\gg [f_{10}^{(\lambda)}]_n \gg 
[f^{(\lambda)}_{1,-1}]_n \gg [f_{20}^{(\lambda)}]_n \cdots
\ . \label{order}
\end{eqnarray}
As a result, neglected terms are expected to be sufficiently small
in the present numerical estimation.

In short, in this subsection, we have constructed the Borel
transforms of $\Phi_0, \ \Psi_0$ following the procedure discussed
in the previous subsection. The result is summarized as
\begin{eqnarray}
V_{0}(p)&=&4\pi^{3}A_{1}f_{11}^{(I)}(p)+\pi A_{2}f^{(I)}_{1,-1}(p)
\nonumber
\\
U_{0}(p)&=&\pm\left(
4\pi^{3}A_{1}f_{11}^{(I)}(p)+2\pi^{3}A_{1}f_{10}^{(I)}(p)
+\pi A_{2}f^{(I)}_{1,-1}(p)
\right)
\label{sol:numerical0}
\end{eqnarray}

\subsection{Borel Transforms of $\Phi_{01}$ and $\Psi_{01}$}

Let $\Phi_{0j}^-(z),\ \Psi_{0j}^-(z)$ be the sum of negative powers of $z$ in the asymptotic expansion of $\Phi_{0j}(z),\ \Psi_{0j}(z)$ and 
$V_j(p), U_j(p)$ be the Borel transform of $\Phi^-_{0j},\ \Psi^-_{0j}$ 
in the sector including $z=-\infty$, then
$V_j(p)$ provides the following terms in another sector
including $z=+\infty$
\begin{eqnarray}
-\int_{\gamma} dp \hspace*{1mm} e^{-pz} \hspace*{1mm}
V_{j}(p)
=\sum_{n=1}^{\infty}\Phi_{nj}(z) e^{-2\pi inz}
\end{eqnarray}
In addition, we define $\Phi_{n}(z) =\sig{j}\sigma^{j}\Phi_{nj}(z)$
for later use.
Similar quantities $\Psi_{nj}$ and $\Psi_n$ are introduced for
$\Psi$.

As the construction of $V_0, \ U_0$ from $\Phi_{10}, \ \Psi_{10}$,
$V_1, \ U_1$ can be obtained from the solutions $\Phi_{11}, \ \Psi_{11}$
of the following equations:
\begin{eqnarray}
\Delta \Phi_{11}(z)&=&
\pm\frac{\Psi_{11}(z)\pm\frac{k-1}{2}z\Psi_{10}(z)}{z}e^{\mp i\Psi_{00}(z)}
-i\frac{\Psi_{10}(z)\Psi_{01}(z)}{z}e^{\mp i\Psi_{00}(z)}
\nonumber
\\
\Delta \Psi_{11}(z-1)&=&
\pm\frac{\Phi_{11}(z)\pm\frac{1-k}{2}z\Phi_{10}(z)}{z}e^{i\Phi_{00}(z)}
\pm i\frac{\Phi_{10}(z)\Phi_{01}(z)}{z}e^{i\Phi_{00}(z)}
\label{eq:11}
\end{eqnarray}
Indeed, from this equation, asymptotic expansions of $\Phi_{11}, \ \Psi_{11}$
in $z$ are obtained and, by exactly the same procedure as in the 
previous subsection, $V_1, \ U_1$ are determined as (for more detail,
see \ref{app01})
\begin{eqnarray}
V_{1}(p)&=&
\pm\left(
8\pi^{4}B_{2}(k-1)f_{12}^{(I)}(p)+
4\pi^{3}(B_{3}(k+1)\mp B_{1}kt_{c})f_{11}^{(R)}(p)
\right)
\nonumber
\\
U_{1}(p)&=&-8\pi^{4}B_{2}(k-1)f_{12}^{(I)}(p)+
4\pi^{3}(B_{3}(k+1)\mp B_{1}kt_{c})f_{11}^{(R)}(p)
\nonumber
\\
&&\mskip 270 mu
-4\pi^{3}
B_{4}(k-1)f_{11}^{(I)}(p)
\label{sol:numerical1}
\end{eqnarray}
And $B_1$, $B_2$, $B_3$
and $B_4$ can be numerically evaluated as $A_1$, $A_2$ and $A_3$:
\begin{eqnarray}
B_1=0.7303,\ 
B_2=0.007399,\ 
B_3=0.03651,\ 
B_4=0.04649 \ . 
\label{Bs}
\end{eqnarray}

\subsection{General Structure of Inner Solution and Stokes Multiplier}

In this subsection, we study the general structure of the inner solution
$\Phi_1(z)\equiv \sum_{j=0}^\infty \sigma^j \Phi_{1j}(z)$  
$\Psi_1(z)\equiv \sum_{j=0}^\infty \sigma^j \Psi_{1j}(z)$, 
which satisfies the following equation:
\begin{eqnarray}
\Delta \Phi_1(z)&=& -\sigma \Psi_1(z)\cos\left(\Psi_0(z)+u_{00}(t_c+\sigma z)\right)
\nonumber \\
\Delta \Psi_1(z-1)&=& k\sigma \Phi_1(z)\cos\left(\Phi_0(z)+v_{00}(t_c+\sigma z)\right)
\label{phi1}
\end{eqnarray}
For this purpose, it is convenient to introduce two solutions of the above equation
admitting the expansions:
\begin{eqnarray}
\left[\matrix{\Phi_{10}^{(A)}(z) \cr \Psi_{10}^{(A)}(z)}\right] 
+ \sum_{j=1}^\infty \sigma^j \left[\matrix{\Phi_{1j}^{(A)}(z) \cr \Psi_{1j}^{(A)}(z)}\right] 
\ , \qquad
\left[\matrix{\Phi_{10}^{(B)}(z) \cr \Psi_{10}^{(B)}(z)}\right] 
+ \sum_{j=1}^\infty \sigma^j \left[\matrix{\Phi_{1j}^{(B)}(z) \cr \Psi_{1j}^{(B)}(z)}\right] 
\end{eqnarray}
where $\Phi_{1j}^{(A)}(z)$ and $\Phi_{1j}^{(B)}(z)$ are uniquely determined by posing
$$
\left\{{\rm coefficient \ of \ } {1\over z} {\rm \ in \ } \Phi_{1j}^{(A)}(z)\right\} = 
\left\{{\rm coefficient \ of \ } z {\rm \ in \ } \Phi_{1j}^{(B)}(z)\right\} = 0 \ .
$$

Then, because of the linearity of (\ref{phi1}), one has the following expressions
\begin{eqnarray}
\Phi_{10}&=&{\widetilde\Lambda}_0\Phi^{(A)}_{10}+\Lambda_0\Phi^{(B)}_{10}
\nonumber
\\
\Phi_{11}&=&{\widetilde\Lambda}_0\Phi^{(A)}_{11}+\Lambda_0\Phi^{(B)}_{11}+
{\widetilde\Lambda}_1\Phi^{(A)}_{10}+\Lambda_1\Phi^{(B)}_{10}
\nonumber
\\
&&\cdots\cdots\cdots
\nonumber
\\
\Phi_{1n}&=&\sum_{j=0}^{n}{\widetilde\Lambda}_k \Phi_{1,n-k}^{(A)}
+\sum_{j=0}^{n}\Lambda_k \Phi_{1,n-k}^{(B)}
\end{eqnarray}
which leads to
\begin{eqnarray}
\Phi_1(z)=({\widetilde\Lambda}_0&+&\sigma {\widetilde\Lambda}_1+\cdots)
\left(\Phi^{(A)}_{10}+\sigma\Phi^{(A)}_{11}+\cdots\right)
\nonumber
\\
&+&
(\Lambda_0+\sigma \Lambda_1+\cdots)
\left(\Phi^{(B)}_{10}+\sigma\Phi^{(B)}_{11}+\cdots\right) \ .
\end{eqnarray}
The coefficient $\displaystyle \Lambda=\sig{n}\sigma^n \Lambda_n$
is referred to as the Stokes multiplier in 
\cite{NakamuraKushibe}. Let $\displaystyle \widetilde\Lambda=\sig{n}
\sigma^n \widetilde\Lambda_n$, then one finally has
$$
\left[\matrix{\Phi_1 \cr \Psi_1}\right]=\widetilde\Lambda 
\left[\matrix{ \Phi_{10}^{(A)} +\sigma \Phi_{11}^{(A)}+ \cdots \cr 
\Psi_{10}^{(A)} +\sigma \Psi_{11}^{(A)} +\cdots }\right]
+ \Lambda \left[\matrix{\Phi_{10}^{(B)} +\sigma \Phi_{11}^{(B)} + \cdots  \cr 
\Psi_{10}^{(B)} +\sigma \Psi_{11}^{(B)} +\cdots }\right] \ .
$$

The first two terms of the Stokes multiplier can be estimated as follows:
\begin{eqnarray}
&&
\left[
\begin{array}{l}
\displaystyle -\int_{\gamma} dp \hspace*{1mm} e^{-pz} \hspace*{1mm}
\{V_0(p) + \sigma V_1(p)\}
\\
\displaystyle -\int_{\gamma} dp \hspace*{1mm} e^{-pz} \hspace*{1mm}
\{U_0(p) + \sigma U_1(p)\}
\end{array}
\right]
\nonumber
\\
\nonumber
=&&\left(
\Lambda
\left[
\begin{array}{l}
(\Phi_{10}^{(B)}+\Phi_{11}^{(B)}\sigma+\cdots)
\\ 
(\Psi_{10}^{(B)}+\Psi^{(B)}_{11}\sigma+\cdots)
\end{array}
\right]
+
\widetilde\Lambda
\left[
\begin{array}{l}
(\Phi_{10}^{(A)}+\Phi_{11}^{(A)}\sigma+\cdots)
\\ 
(\Psi_{10}^{(A)}+\Psi^{(A)}_{11}\sigma+\cdots)
\end{array}
\right]
\right)
e^{-2\pi iz}
\nonumber
\\
=
&&\left[
\begin{array}{l}
\pm\Lambda_{0}z+\sigma (\pm\Lambda_{1}z \pm \frac{k-1}{6}\Lambda_{0}z^{2})
\\
\Lambda_{0}z+\sigma\left(\Lambda_{1}z - \frac{k-1}{6}\Lambda_{0}(z^{2}+z)\right)\end{array}
\right]
e^{-2\pi iz}
\label{matching}
\end{eqnarray}
where the contour integral is evaluated with the aid of 
(\ref{sol:numerical0}) and (\ref{sol:numerical1}).
This leads to the following relations
\begin{eqnarray}
\Lambda_{0}&=&i4\pi^{3}A_{1}=i48\pi^{4}B_{2}=i24\pi^{3}B_{4}
\nonumber
\\
\Lambda_{1}&=&4\pi^{3}(B_{3}(k+1)\mp B_{1}kt_{c})
\label{StokesVal}
\end{eqnarray}
Eq.(\ref{StokesVal}) implies two linear relations among
$A_1$, $B_2$ and $B_4$, which are satisfied by the present
numerical estimations rather well:
\begin{eqnarray}
{A_1\over 12 \pi B_2}=0.99998\simeq 1 \qquad
{A_1\over 6 \pi B_4}=0.99997\simeq 1 \nonumber \ .
\end{eqnarray}

\section{Matching of Inner and Outer Solutions}

\subsection{Matching at a Singular Point}
In this section, solutions of the outer equations are constructed
and are matched with the inner solutions.
We first consider the contribution from the singularity $t_1$.
Corresponding to the expansion of the analytically continued inner solutions:
$\Phi=\sum_{n=0}^\infty \Phi_n e^{-2\pi i nz}$, $\Psi=\sum_{n=0}^\infty \Psi_n e^{-2\pi i nz}$, 
the original solutions $v$ and $u$ acquire new terms in a sector ${\rm Re} \ t\ge{\rm Re}\ t_1$ of the 
$t_1$-neighborhood:
\begin{eqnarray}
v(t)=v_{0}(t,\sigma)+v_{1}(t,\sigma)e^{-\frac{2\pi i}{\sigma}t}+v_{2}(t,\sigma)e^{-\frac{4\pi i}{\sigma}t}+ \cdots 
\nonumber
\\
u(t)=u_{0}(t,\sigma)+u_{1}(t,\sigma)e^{-\frac{2\pi i}{\sigma}t}+u_{2}(t,\sigma)e^{-\frac{4\pi i}{\sigma}t}+ \cdots
\label{original new exp}
\end{eqnarray}
Because of $e^{-2\pi i n z}\propto \epsilon^n e^{-{2\pi i n \over \sigma}t}$ with $\epsilon \equiv e^{-{\pi^2 \over \sigma \sqrt{k}}}$,
$v_n$ and $u_n$ are of order of $\epsilon^n$.
By substituting  (\ref{original new exp}) into 
(\ref{eq:Harper}) and comparing term by term,
we obtain the equations for $v_1$ and $u_1$:
\begin{eqnarray}
v_{1}(t+\sigma)-v_{1}(t) &=&-\sigma u_{1}(t) \cos u_{0}(t)
\nonumber
\\
u_{1}(t+\sigma)-u_{1}(t) &=&k\sigma v_{1}(t+\sigma) \cos v_{0}(t+\sigma)
\label{Harper1}
\end{eqnarray}

Its solution is uniquely determined by the matching condition:
\begin{eqnarray}
\left. e^{-\frac{2\pi it}{\sigma}}
\left[
\matrix{
v_{1}(t,\sigma)
\cr
u_{1}(t,\sigma)
}
\right]\right|_{t=t_1+\sigma z}
&=& e^{-2\pi iz}\biggl\{
\widetilde\Lambda 
\left[\matrix{ \Phi_{10}^{(A)}(z) +\sigma \Phi_{11}^{(A)}(z)+ \cdots \cr 
\Psi_{10}^{(A)}(z) +\sigma \Psi_{11}^{(A)}(z) +\cdots }\right]
\nonumber \\
&&\mskip 50mu+ \Lambda \left[\matrix{\Phi_{10}^{(B)}(z) +\sigma \Phi_{11}^{(B)}(z) + \cdots  \cr 
\Psi_{10}^{(B)}(z) +\sigma \Psi_{11}^{(B)}(z) +\cdots }\right] \biggl\} \ ,
\end{eqnarray}
or equivalently
\begin{eqnarray}
\left[
\matrix{
v_{1}(t_1+\sigma z,\sigma)
\cr
u_{1}(t_1+\sigma z,\sigma)
}
\right]
&=& e^{\frac{2\pi it_1}{\sigma}}
 \biggl\{
\widetilde\Lambda 
\left[\matrix{ \Phi_{10}^{(A)}(z) +\sigma \Phi_{11}^{(A)}(z)+ \cdots \cr 
\Psi_{10}^{(A)}(z) +\sigma \Psi_{11}^{(A)}(z) +\cdots }\right]
\nonumber \\
&&\mskip 90mu+ \Lambda \left[\matrix{\Phi_{10}^{(B)}(z) +\sigma \Phi_{11}^{(B)}(z) + \cdots  \cr 
\Psi_{10}^{(B)}(z) +\sigma \Psi_{11}^{(B)}(z) +\cdots }\right] \biggl\} \ .
\label{matchrel}
\end{eqnarray}

The matching condition suggests the following expansions:
\begin{eqnarray}
v_1(t,\sigma)=\sigma^j e^{\frac{2\pi it_1}{\sigma}}\sig{n}v_{1n}(t)\sigma^n
\ , \qquad
u_1(t,\sigma)=\sigma^j e^{\frac{2\pi it_1}{\sigma}}\sig{n}u_{1n}(t)\sigma^n
\end{eqnarray}
where $j$ is an integer. Then equations for $v_{10},u_{10}\cdots$ are given by
\begin{eqnarray}
&&v'_{10}(t)=-u_{10}(t)\cos u_{00}(t)
\nonumber
\\
&&u'_{10}(t)=kv_{10}(t)\cos v_{00}(t)
\label{ODE:10}
\\
&&v'_{11}(t)+\frac{1}{2}v''_{10}(t)=-u_{11}(t)\cos u_{00}(t)+u_{10}(t)u_{01}(t)\sin u_{00}(t)
\nonumber
\\
&&u'_{11}(t)-\frac{1}{2}u''_{10}(t)=k\left(v_{11}(t)\cos v_{00}(t)-v_{10}(t)v_{01}(t)\sin v_{00}(t)\right)
\label{ODE:11}
\end{eqnarray}

First, we solve (\ref{ODE:10}). Since it admits two linearly independent solutions: 
$\left[\matrix{x_1 \cr y_1}\right]$ of (\ref{eq:x1}) and the following
$\left[\matrix{x_2 \cr y_2}\right]$:
\begin{eqnarray}
x_{2}(t)&=&\frac{-x_{1}(t)}{4k(1-k)}
\left[\frac{(1-k)^{2}}{2\sqrt{k}}
\left(\sh{t}\ch{t}+\sqrt{k}t\right)
-k^{3/2}\tanh\sqrt{k}t
\right]
\nonumber
\\
y_{2}(t)&=&\frac{y_{1}(t)}{4k(1-k)}
\left[\frac{(1-k)^{2}}{2\sqrt{k}}
\left(\sh{t}\ch{t}-\sqrt{k}t\right)
+\frac{1}{\sqrt{k}}\coth\sqrt{k}t
\right] \ ,
\nonumber
\\
\label{eq:x2}
\end{eqnarray}
one has
\begin{eqnarray}
\left[
\matrix{
v_{10}(t)
\cr
u_{10}(t)
}
\right]
&=&
a\left[
\matrix{
x_1(t)
\cr	
y_1(t)
}
\right]
+
b\left[
\matrix{
x_2(t)
\cr	
y_2(t)
}
\right]
\end{eqnarray}
with the matching condition
\begin{eqnarray}
\sigma^j
\left.
\left[
\matrix{
v_{10}(\sigma z +t_1)
\cr
u_{10}(\sigma z +t_1)
}
\right]
\right|_{\sigma^0}
&=&
\sigma^j 
\left.
\left\{
a\left[
\matrix{
x_1(\sigma z +t_1)
\cr	
y_1(\sigma z +t_1)
}
\right]
+
b\left[
\matrix{
x_2(\sigma z +t_1)
\cr	
y_2(\sigma z +t_1)
}
\right]
\right\}
\right|_{\sigma^0}
\nonumber \\
&=&
\left.
\widetilde\Lambda_0
\left[
\matrix{
\Phi^{(A)}_{10}(z)
\cr	
\Psi^{(A)}_{10}(z)
}
\right]
+
\Lambda_0 \left[
\matrix{
\Phi^{(B)}_{10}(z)
\cr	
\Psi^{(B)}_{10}(z)
}
\right]
\right|_{\rm dom.}
\nonumber
\\
&=&
\Lambda_0
\left[
\matrix{
z
\cr	
z
}
\right]
\label{apprMatch}
\end{eqnarray} 
where the subscript $\sigma^0$ indicates to take the terms of order $\sigma^0$ 
and dom. stands for the largest part for $z\to \infty$.

Because L.H.S. of (\ref{apprMatch}) starts from $z$, one should have $j=-1$.
Then, the coefficients $a$ and $b$ are determined by the requirement that
(\ref{apprMatch}) admits well defined $\sigma\to 0$ limit:
$$
a=\Lambda_0
\frac{t_{1}(k-1)^{2} + (1+k)}{4ik(k-1)}
\ , \qquad
b=-\frac{2\Lambda_0}{i} \ .
$$
Therefore, up to the $0$th order in $\sigma$, we have
\begin{eqnarray}
\left[
\begin{array}{l}
v_{1}(t)
\\ 
u_{1}(t)
\end{array}
\right]
\approx
\frac{2\Lambda_0}{i\sigma}
e^{\frac{2\pi it_{1}}{\sigma}}
\left\{
\frac{t_{1}(k-1)^{2} + (1+k)}{8k(k-1)}
\left[
\begin{array}{l}
x_{1}(t)
\\ 
y_{1}(t)
\end{array}
\right]
-
\left[
\begin{array}{l}
x_{2}(t)
\\ 
y_{2}(t)
\end{array}
\right]
\right\}
\label{ac}
\end{eqnarray}
Similarly, up to the first order in $\sigma$, one obtains
\begin{eqnarray}
\left[
\begin{array}{l}
v_{1}(t)
\\ 
u_{1}(t)
\end{array}
\right]
&\approx&
\frac{2\Lambda^{(1)}}{i\sigma}
e^{\frac{2\pi it_{1}}{\sigma}}
\Biggl\{
\frac{t_{1}(k-1)^{2} + (1+k)}{8k(k-1)}
\left[
\begin{array}{l}
x_{1}(t)
\\ 
y_{1}(t)+\sigma y'_{1}(t)/2
\end{array}
\right]
\nonumber
\\
&&\mskip 240 mu-
\left[
\begin{array}{l}
x_{2}(t)
\\ 
y_{2}(t)+\sigma y'_{2}(t)/2
\end{array}
\right]
\Biggr\}
\label{af}
\end{eqnarray}
where $\Lambda^{(1)}\equiv\Lambda_0+\sigma\Lambda_1$ is the first order Stokes
multiplier.

In the $t_2$-neighborhood, the unperturbed solutions $v_0$, $u_0$ acquire
the following terms in a sector ${\rm Re} \ t\ge{\rm Re}\ t_2$:
\begin{eqnarray}
\left[
\begin{array}{l}
v_{1}(t)
\\ 
u_{1}(t)
\end{array}
\right]
&\approx&
\frac{2\Lambda^{(2)}}{i\sigma}
e^{\frac{2\pi it_{2}}{\sigma}}
\Biggl\{
\frac{t_{2}(k-1)^{2} - (1+k)}{8k(k-1)}
\left[
\begin{array}{l}
x_{1}(t)
\\ 
y_{1}(t)+\sigma y'_{1}(t)/2
\end{array}
\right]
\nonumber \\
&& \mskip 240 mu -
\left[
\begin{array}{l}
x_{2}(t)
\\ 
y_{2}(t)+\sigma y'_{2}(t)/2
\end{array}
\right]
\Biggr\}
\label{second}
\end{eqnarray}
where $\Lambda^{(2)}$ is the first order Stokes multiplier arising from
the singularity $t_2$. 

\subsection{Contributions from Other Singular Points}
So far, contributions from $t_1$ and $t_2$ have been considered.
Here, we study contributions from other singular points.

Let  $t^{(n)}$ denote the singular points satisfying, 
$|{\rm Im}[t^{(n)}]| = \frac{(2n+1)\pi^2}{2\sqrt{k}}$.
Then terms arising from $t^{(n)}$ are of order $\epsilon^{2n+1}$, 
where $\epsilon=e^{-{\pi^2\over \sqrt{k}\sigma}}$. 
Since $(v_0(t),u_0(t))$ is bounded and 
$(v_n(t),u_n(t))\approx e^{\sqrt{k}t},\ (n\ge 1)$ for sufficiently large $t$,
one can neglect the terms arising from $t^{(n)}$ ($n\ge 1$)
even for sufficiently large $t$.

On the other hand, terms arising from $t_1$ may produce new terms when $t$ passes through 
$t_2$-neighborhood, but they 
are negligible as they are higher order with respect to $\epsilon$. Therefore, the 
overall contribution 
of the singular points is a simple sum of the contributions from 
$t_1,\ t_2,\ t_1^*,\ t_2^*$ and one
has the following solution of the unstable manifold in the whole 
time domain:
\begin{eqnarray}
v_{u}(t)&&=
v_{00}(t)+\sigma^2 v_{02}(t) 
\nonumber \\
&&~+S_-(t)
{\rm Re}\left[
\frac{4\Lambda^{(1)}}{i\sigma}
e^{\frac{2\pi it_{1}}{\sigma}}
\left(\frac{t_{1}(k-1)^{2} + (1+k)}{8k(k-1)}x_{1}(t)-x_{2}(t)\right)
e^{-\frac{2\pi it}{\sigma}}
\right]
\nonumber
\\
&&~+S_+(t)
{\rm Re}\left[
\frac{4\Lambda^{(2)}}{i\sigma}
e^{\frac{2\pi it_{2}}{\sigma}}
\left(\frac{t_{2}(k-1)^{2} - (1+k)}{8k(k-1)}x_{1}(t)-x_{2}(t)\right)
e^{-\frac{2\pi it}{\sigma}}
\right]
\nonumber
\\ \nonumber \\
u_{u}(t)
&&=
u_{00}(t)+\sigma \frac{y_{1}(t)}{2}+\sigma^{2}u_{02}(t)+\sigma^{3} 
\left(\frac{1}{2}u'_{02}(t)-\frac{1}{24}y''_{1}(t)\right)
\nonumber
\\
&&~+S_-(t)
{\rm Re}\biggl[
\frac{4\Lambda^{(1)}}{i\sigma}
e^{\frac{2\pi it_{1}}{\sigma}}
\biggl\{\frac{t_{1}(k-1)^{2} + (1+k)}{8k(k-1)}\left(y_{1}(t)+\sigma\frac{y'_{1}(t)}{2}
\right)\nonumber \\
&&\mskip 270 mu -
\left(y_{2}(t)+\sigma\frac{y'_{2}(t)}{2}\right)\biggr\}
e^{-\frac{2\pi it}{\sigma}}
\biggr]
\nonumber
\\
&&~+S_+(t)
{\rm Re}\biggl[
\frac{4\Lambda^{(2)}}{i\sigma}
e^{\frac{2\pi it_{2}}{\sigma}}
\biggl\{ \frac{t_{2}(k-1)^{2} - (1+k)}{8k(k-1)}\left(y_{1}(t)+\sigma\frac{y'_{1}(t)}{2}\right)
\nonumber \\
&& \mskip 270 mu -
\left(y_{2}(t)+\sigma\frac{y'_{2}(t)}{2}\right) \biggr\}
e^{-\frac{2\pi it}{\sigma}}
\biggr]
\nonumber
\\ \label{unstableMan}
\end{eqnarray}
where $S_\pm(t)=S(t\pm t_1^{R})$ and $S(t)$ stands for 
the step function, 
and $\Lambda^{(1)}$,\ $\Lambda^{(2)}$ denote the first order Stokes multipliers from 
the analysis of $t_1$,\ $t_2$, respectively.

\subsection{Comparison with Numerical Calculation}
Here we compare the analytical solution (\ref{unstableMan}) obtained in the previous 
subsection with the numerical calculation. For this purpose, it is convenient to 
rewrite the solution as
\begin{eqnarray}
&&v_{u}(t)=\widetilde{v}_{0}(t)
+2S(t)
\Biggl[ a_{11}(t) \cos\frac{2\pi t}{\sigma} + b_{11}(t) \sin\frac{2\pi t}{\sigma}
\Biggr]
\nonumber
\\
&&~~~~~~+2S(t+2t_{1}^{R})
\Biggl[ a_{12}(t) \cos\frac{2\pi(t+2t_{1}^{R})}{\sigma} + b_{12}(t) \sin\frac{2\pi(t+2t_{1}^{R})}{\sigma}
\Biggr]
\nonumber
\\
&&u_{u}(t)=\widetilde{u}_{0}(t)
+2S(t)
\Biggl[ a_{21}(t) \cos\frac{2\pi t}{\sigma} + b_{21}(t) \sin\frac{2\pi t}{\sigma}
\Biggr]
\nonumber
\\
&&~~~~~~+2S(t+2t_{1}^{R})
\Biggl[ a_{22}(t) \cos\frac{2\pi(t+2t_{1}^{R})}{\sigma} + b_{22}(t) \sin\frac{2\pi(t+2t_{1}^{R})}{\sigma}
\Biggr]
\nonumber \\ \label{unstableMan2}
\end{eqnarray}
where the auxiliary functions are given by
\begin{eqnarray}
a_{11}(t)&=&(K+K_1)
\Big(\alpha \widetilde{x}_1(t)-\widetilde{x}_2(t)\Big)
-K'_1 \beta \widetilde{x}_1(t)
\nonumber
\\
b_{11}(t)&=&K'_1
\Big(\alpha \widetilde{x}_1(t)-\widetilde{x}_2(t)\Big)
+(K+K_1)\beta \widetilde{x}_1(t)
\nonumber
\\
a_{12}(t)&=&(K_1 -K)
\Big(\alpha \widetilde{x}_1(t)+\widetilde{x}_2(t)\Big)
-K'_1 \beta \widetilde{x}_1(t)
\nonumber
\\
b_{12}(t)&=&-K'_1
\Big(\alpha \widetilde{x}_1(t)+\widetilde{x}_2(t)\Big)
+
(K-K_1)\beta \widetilde{x}_1(t)
\nonumber
\\
a_{21}(t)&=&(K+K_1)
\left\{
\alpha\left(\widetilde{y}_1(t)+\frac{\sigma}{2}\widetilde{y}'_{1}(t)\right)
-\left(\widetilde{y}_2(t)+\frac{\sigma}{2}\widetilde{y}'_{2}(t)\right)
\right\} \nonumber \\
&&~~~-K'_1 \beta
\left(\widetilde{y}_1(t)+\frac{\sigma}{2}\widetilde{y}'_{1}(t)\right)
\nonumber
\\
b_{21}(t)&=&K'_1
\left\{\alpha 
\left(\widetilde{y}_1(t)+\frac{\sigma}{2}\widetilde{y}'_{1}(t)\right)
-\left(\widetilde{y}_2(t)+\frac{\sigma}{2}\widetilde{y}'_{2}(t)\right)
\right\} \nonumber \\
&&~~~+(K+K_1)\beta 
\left(\widetilde{y}_1(t)+\frac{\sigma}{2}\widetilde{y}'_{1}(t)\right)
\nonumber
\\
a_{22}(t)&=&(K_1 -K)
\left\{
\alpha\left(\widetilde{y}_1(t)+\frac{\sigma}{2}\widetilde{y}'_{1}(t)\right)
-\left(\widetilde{y}_2(t)+\frac{\sigma}{2}\widetilde{y}'_{2}(t)\right)
\right\} \nonumber \\
&&~~~-K'_1 \beta
\left(\widetilde{y}_1(t)+\frac{\sigma}{2}\widetilde{y}'_{1}(t)\right)
\nonumber
\\
b_{22}(t)&=&-K'_1
\left\{\alpha 
\left(\widetilde{y}_1(t)+\frac{\sigma}{2}\widetilde{y}'_{1}(t)\right)
+\left(\widetilde{y}_2(t)+\frac{\sigma}{2}\widetilde{y}'_{2}(t)\right)
\right\} \nonumber \\
&&~~~+(K-K_1)\beta 
\left(\widetilde{y}_1(t)+\frac{\sigma}{2}\widetilde{y}'_{1}(t)\right)
\nonumber
\\
\end{eqnarray}
The functions with tilde are defined as ${\widetilde f}(t)\equiv f(t+t_1^R)$
and the constants $\ \alpha,\ \beta,\ K_{1},\ K'_{1}$ are given by
\begin{eqnarray}
K&=&\frac{8\pi^{3}A_{1}}{\sigma}e^{-\frac{\pi^{2}}{\sigma\sqrt{k}}}
\ , \qquad 
K_{1}=-8\pi^{3}B_{1}t_{1}^{I}e^{-\frac{\pi^{2}}{\sigma\sqrt{k}}}
\nonumber
\\
K'_{1}&=&8\pi^{3}(B_{1}kt_{1}^{R}-B_{3}(k+1))e^{-\frac{\pi^{2}}{\sigma\sqrt{k}}}\nonumber
\\
\alpha&=&
\frac{t_{1}^{R}(k-1)^{2}+(1+k)}{8k(k-1)}
\ ,
\qquad
\beta=\frac{t_{1}^{I}(k-1)}{8k}
\end{eqnarray}
where $t_{1}^{R}$ and $t_{1}^{I}$ are the real and imaginary
parts of $t_1$, respectively.
Eq.(\ref{unstableMan2}) indicates that, when $t$ exceeds $t=0$, exponentially
growing oscillatory term appears and, when $t$ exceeds $t=-2t^{R}_1 (>0)$,
another exponentially growing term is added. These terms should describe heteroclinic tangles.

\begin{figure}[htbp]
\begin{center}
\includegraphics[width=8cm,keepaspectratio,clip]{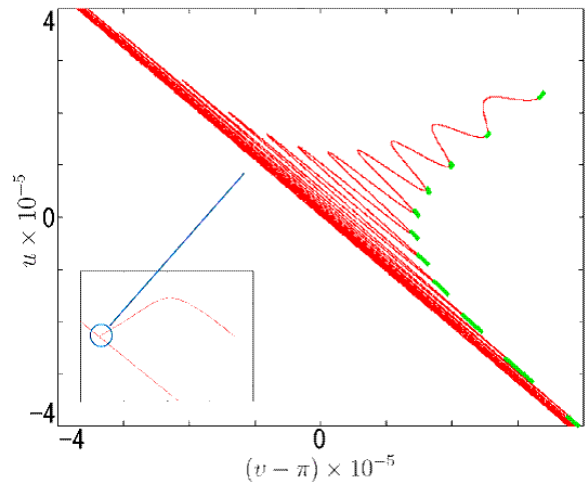}
\caption{The analytically constructed unstable manifold (solid line)
near $(-\pi,0)$ and 
time evolution of the ensemble (dots).
Inset shows the overall view of the analytically construced unstbale manifold.\ $\sigma=0.35,k=0.85$}
\label{fig:globaal}
\end{center}
\end{figure}

\begin{figure}[htbp]
\begin{center}
\includegraphics[width=8cm,keepaspectratio,clip]{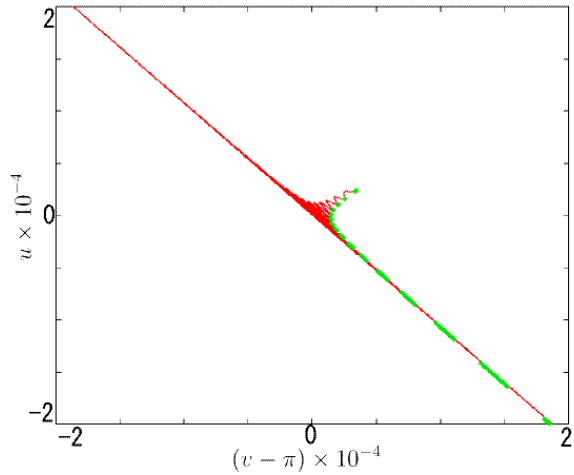}
\caption{The analytically constructed unstable manifold (solid line) 
near $(-\pi,0)$ and 
time evolution of the ensemble (dots).\ $\sigma=0.35,k=0.85$}
\label{fig:result1}
\end{center}
\end{figure}

\begin{figure}[htbp]
\begin{center}
\includegraphics[width=8cm,keepaspectratio,clip]{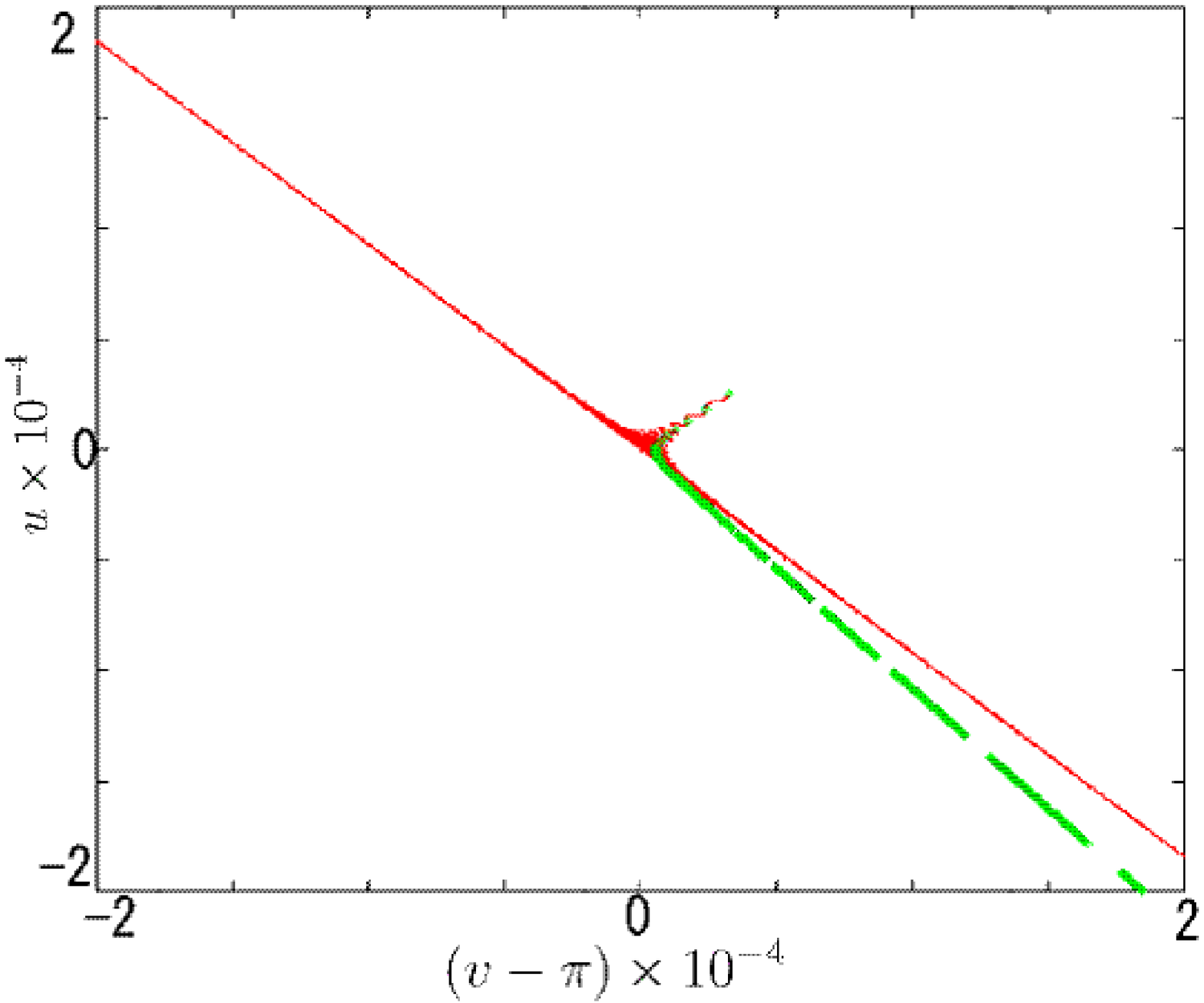}
\caption{The analytically constructed unstable manifold (solid line) 
near $(-\pi,0)$ and 
time evolution of the ensemble (dots).\ Only the leading term of the inner equation is taken into consideration.\ $k=0.85,\sigma=0.35$}
\label{fig:result2}
\end{center}
\end{figure}

This is indeed the case. Fig.~\ref{fig:globaal},\ Fig.~\ref{fig:result1}
 shows an analytical solution of the unstable manifold from $(\pi,0)$. 
One can see a large-amplitude oscillation near $(-\pi,0)$.
As shown in Fig.~\ref{fig:globaal},\ Fig.~\ref{fig:result1},
the analytical solution (\ref{unstableMan2}) (solid curve) well reproduces the 
oscillation in the numerically calculated orbit (dots). 
The dots show the time evolution of an ensemble consisting of 500 points.
One can see the stretching of the ensemble is also well reproduced by the approximate unstable
manifold. When only the leading order terms of the inner equation are retained, the agreement 
between the analytical and numerical 
results is not so good (cf. Fig.~\ref{fig:result2}) and, thus, the terms in the inner solution
of order $\sigma$ are important.

Before closing this section, we give an approximate stable manifold
$(v_s(t),u_s(t))$.
By the similar approach to construct the unstable manifold, 
we have the similar expressions:
\begin{eqnarray}
v_s(t)=&&\widetilde{v}_{0}(t)
-2S(-t)
\Biggl[ a_{11}(t) \cos\frac{2\pi t}{\sigma} + b_{11}(t) \sin\frac{2\pi t}{\sigma}
\Biggr]
\nonumber
\\
&&-2S(-t+2t_{1}^{R})
\Biggl[ a_{12}(t) \cos\frac{2\pi(t+2t_{1}^{R})}{\sigma} + b_{12}(t) \sin\frac{2\pi(t+2t_{1}^{R})}{\sigma}
\Biggr]
\nonumber
\\
u_s(t)=&&\widetilde{u}_{0}(t)
-2S(-t)
\Biggl[ a_{21}(t) \cos\frac{2\pi t}{\sigma} + b_{21}(t) \sin\frac{2\pi t}{\sigma}
\Biggr]
\nonumber
\\
&&-2S(-t+2t_{1}^{R})
\Biggl[ a_{22}(t) \cos\frac{2\pi(t+2t_{1}^{R})}{\sigma} + b_{22}(t) \sin\frac{2\pi(t+2t_{1}^{R})}{\sigma}
\Biggr]
\nonumber \\ \label{stableMan2}
\end{eqnarray}

\section{Reconnection of Unstable Manifold}
We remind that the separatrix of the continuous-limit
equation is the unperturbed stable/unstable 
manifold.
As mentioned in Introduction (cf.Fig.~\ref{fig:differential}), 
the separatrix changes its topology depending on the parameter $k$
(reconnection). When $0< k <1$, there appears a separatrix connecting
$(\pi,0)$ and $(-\pi,0)$. As $k\to 1$, it approaches the union of
two segments: the one connecting $(\pi,0)$ with $(0,\pi)$ and the
other $(0,\pi)$ with $(-\pi,0)$. Both segments are separatrices
at the parameter value $k=1$. 
The corresponding topological change occurs for stable/unstable manifolds,
which will be discussed in this section.
As shown in the previous section, since the approximate stable and 
unstable manifolds are related with each other by simple symmetry,
it is enough to discuss the topological change of the unstable
manifold.

\begin{figure}[tbp]
\begin{center}
\includegraphics[width=8cm,keepaspectratio,clip]{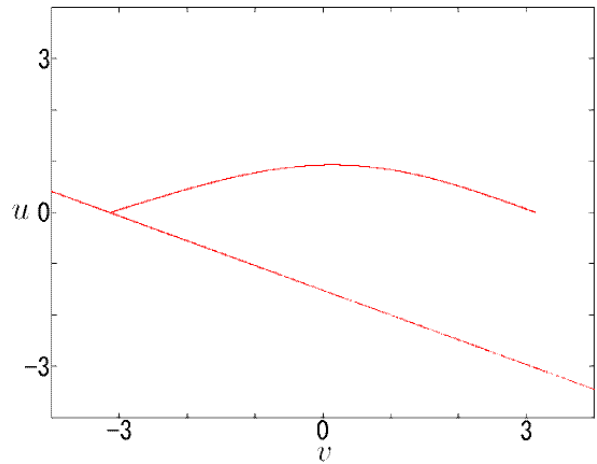}
\caption{The analytically constructed unstable manifold.
 $\sigma=0.35,\ k=0.2$}
\label{0.2}
\end{center}
\end{figure}

\begin{figure}[tbp]
\begin{center}
\includegraphics[width=8cm,keepaspectratio,clip]{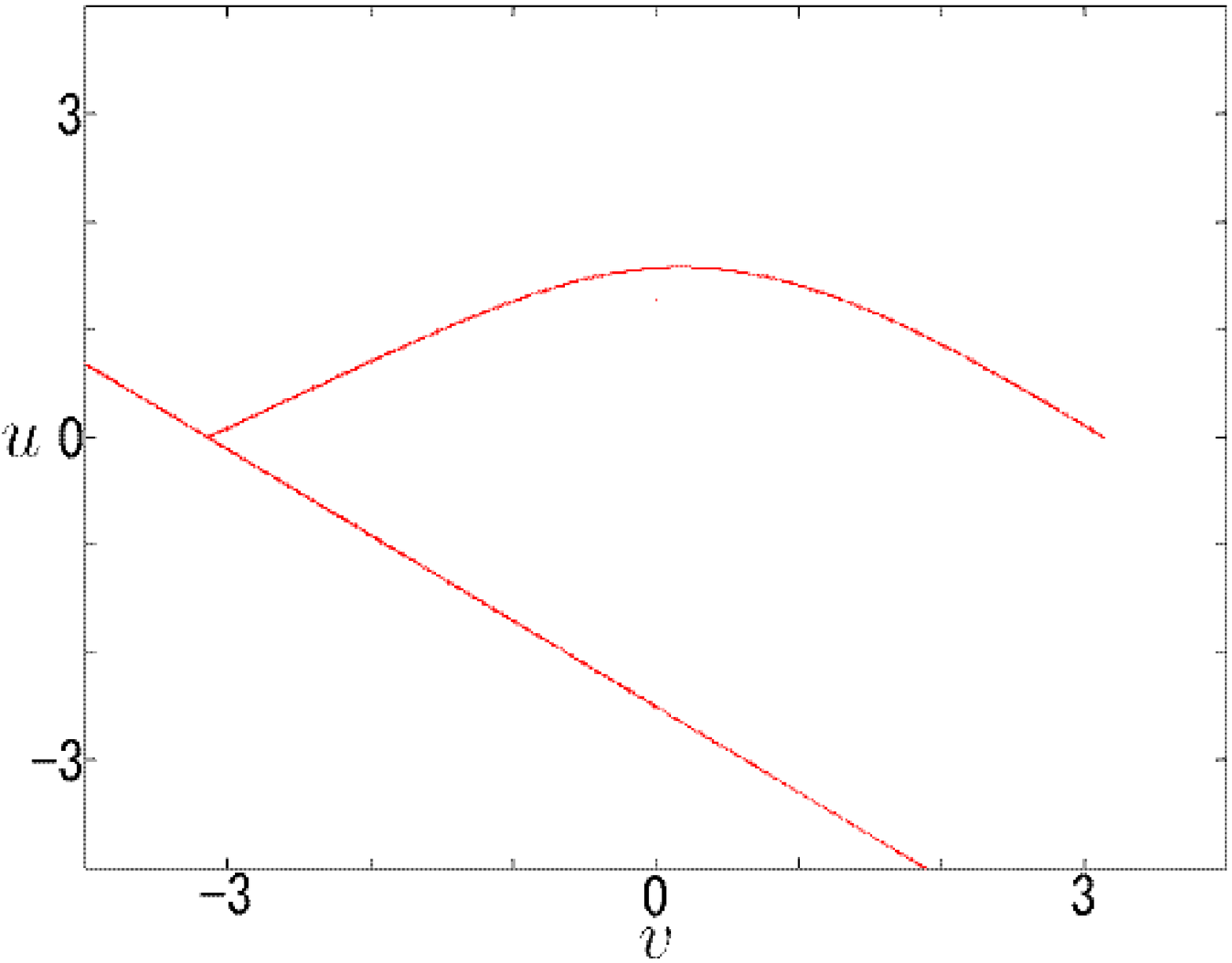}
\caption{The analytically constructed unstable manifold.
 $\sigma=0.35,\ k=0.5$}
\label{0.5}
\end{center}
\end{figure}

\begin{figure}[tbp]
\begin{center}
\includegraphics[width=8cm,keepaspectratio,clip]{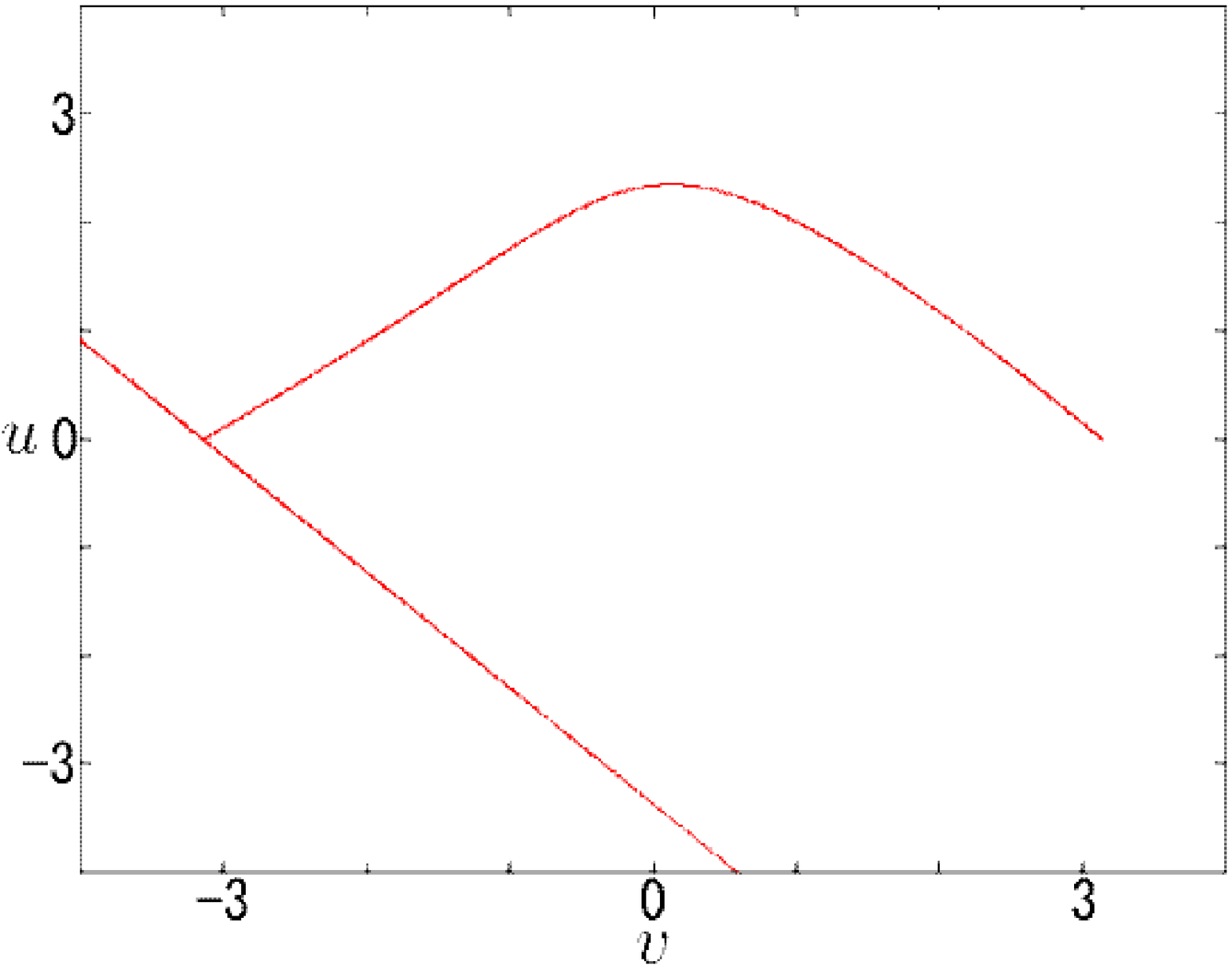}
\caption{Analytically constructed unstable manifold.
 $\sigma=0.35,\ k=0.85$}
\label{0.85}
\end{center}
\end{figure}

\begin{figure}[tbp]
\begin{center}
\includegraphics[width=8cm,keepaspectratio,clip]{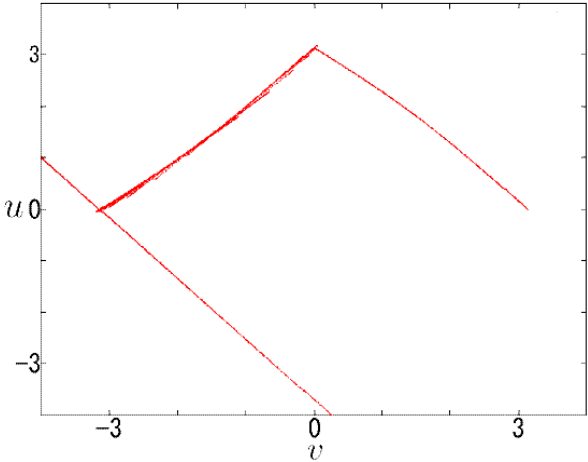}
\caption{The analytically constructed unstable manifold.
 $\sigma=0.35,\ k=1.0-10^{-9}$}
\label{0.99}
\end{center}
\end{figure}

Fig.~\ref{0.2},\ Fig.~\ref{0.5},\ Fig.~\ref{0.85},\ Fig.~\ref{0.99} 
show the unstable manifolds when $\sigma=0.35$ and 
$k=0.2,\ 0.5,\ 0.85,\ 1.0-10^{-9}$, respectively.
The overall structure of the unstable manifold except near the fixed point
$(-\pi,0)$ is very close to that of the separatrix.
Near $(-\pi,0)$, the unstable manifold acquires an oscillatory portion
and the slope of this portion becomes steeper as $k\to 1$.
The behavior of the slope can be understood from the asymptotic ratio 
between the additional terms:
$$
\displaystyle \lim_{t\to +\infty}\frac{u_{1}(t)}{v_{1}(t)+\pi}=-\sqrt{k}-\frac{k}{2}\sigma
\ .
$$

\begin{figure}[t]\label{fig:reconnection}
\begin{center}
\includegraphics[width=10cm,keepaspectratio,clip]{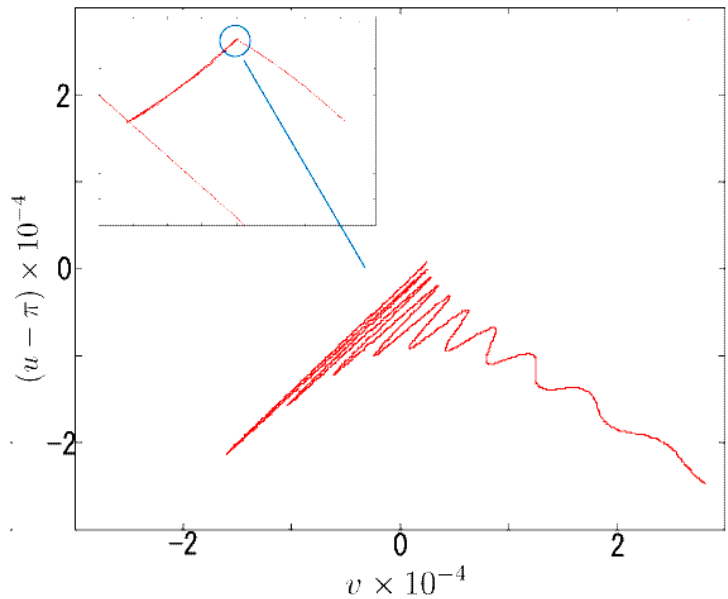}
\caption{The analytically constructed unstable manifold and the oscillation near $(0,\pi)$.\ $\sigma=0.35,\ k=1.0-10^{-9}$}
\end{center}
\end{figure}

It is interesting to see that, when $k$ is very close to unity (i.e., $k=1.0-10^{-9}$), 
there appears 
an additional oscillatory portion in the unstable manifold near $(0,\pi)$
(cf. Fig.~\ref{fig:reconnection}).
As mentioned before, at $k=1$, two segments from ($\pi,0$) to ($0,\pi$) and 
from ($0,\pi$) to ($-\pi,0$)
are separatrices of the continuous-limit equation.
Thus, the perturbed unstable manifold at $k=1$ starting from $(\pi,0)$ should have an
oscillatory portion near $(0,\pi)$. The oscillation near $(0,\pi)$ of the unstable manifold 
shown in Fig.~\ref{fig:reconnection} can be considered as the precursor of the oscillation
in the unstable manifold at $k=1$.

The origin of these behaviors are summarized as follows.
When $k$ is not very close to 1, splitting term is exponentially small compared to the 
unperturbed solution and can be negligible for not too large $t$ .
But this term becomes dominant for sufficiently large $t$ because $x_{2},\ y_{2} \rightarrow \infty$ 
as $t \rightarrow \infty$.
Thus, a large-amplitude oscillation appears near $(-\pi,0)$.
This is not the case when $k$ is very close to 1 because exponentially small terms with respect to 
$\sigma$ are
proportional to $\frac{1}{1-k}$.
Therefore, the additional terms are not too small compared to the unperturbed solution near 
$(0,\pi)$ and we can observe the oscillation near $(0,\pi)$.
We further remark that two oscillatory portions in the unstable manifold
for $k\simeq 1$ come from two singularities $t_1$ and $t_2$ in the complex time plane.
In other words, the two sequences of singularities are necessary to exhibit
the reconnection of stable/unstable manifolds.

\section{Conclusions}

With the aid of ABAO (the asymptotics beyond all orders) method, we have derived 
analytical approximations of the stable/unstable manifolds, which agree rather
well with those obtained by the numerical iteration. 
When $k$ is not close to 1, the perturbed unstable manifold starting from $(\pi,0)$ exhibits 
highly oscillatory behavior near the fixed point ($-\pi,0$) and its overall structure far from
($-\pi,0$) is almost the same as that of the unperturbed unstable manifold.
However, when $k$ is very close to 1, we can observe a oscillation near $(0,\pi)$, which is 
considered to be the precursor of the heteroclinic tangle in the unstable manifold at $k=1$.
In this way, even when the heteroclinic tangle exists, the unstable manifold smoothly
changes its topology as the change of the parameter $k$.

Contrary to the systems studied so far by ABAO method, the unperturbed solution of the Harper map
have two sequences of singular points in the complex time plane and the interference of the
contributions from them might be expected. In this paper, the approximate stable/unstable
manifolds are constructed just adding the contributions from the two sequences, but 
they agree quite well with the numerically constructed unstable manifold.
In a higher order approximation, we need to take into account this interference.
This aspect will be discussed elsewhere.

\section*{Acknowledgments}

The authors thank Hidekazu Tanaka and  Satoshi Nakayama for their contributions
to the early stage of this work. Also, they are grateful to Prof. K. Nakamura, 
Prof. A. Shudo, Dr.S. Shinohara, T. Miyaguchi and Dr. G. Kimura for 
fruitful discussions and useful comments. Particularly, they thank Prof.
Nakamura for turning their attention to Ref.\cite{NakamuraKushibe}.
This work is partially supported by a Grant-in-Aid for Scientific Research
of Priority Areas ``Control of Molecules in Intense Laser Fields'' and the
21st Century COE Program at Waseda University ``Holistic Research and Education
Center for Physics of Self-organization Systems'' both from Ministry of 
Education, Culture, Sports and Technology of Japan, as well as by
Waseda University Grant for Special Research
Projects, Individual Research (2004A-161).

\appendix

\section{Case of $k>1$}\label{appC}
In this section, we consider the case of $k>1$.
Put
\begin{eqnarray}
\tau\equiv kt,\ \widetilde{\sigma}\equiv k\sigma ,
\quad
\widetilde{u}(\tau)\equiv v\left(-\frac{\tau}{k}+\sigma\right),\ 
\widetilde{v}(\tau)\equiv u\left(-\frac{\tau}{k}+\sigma\right)
\label{trans}
\end{eqnarray}
then $\widetilde{v}$, $\widetilde{u}$ are the solution of the 
Harper map with parameter $1/k$:
\begin{eqnarray}
\widetilde{v}(\tau+\widetilde{\sigma})-\widetilde{v}(\tau)&=&
-\widetilde{\sigma} \sin \widetilde{u}(\tau)
\nonumber
\\
\widetilde{u}(\tau+\widetilde{\sigma})-\widetilde{u}(\tau)&=&
{\widetilde{\sigma}\over k} \sin \widetilde{v}(\tau+\widetilde{\sigma})
\end{eqnarray}

\section{Borel-transformed first order solution of inner equation \label{app01}}
The aim of this appendix is to calculate the Borel transforms 
of the solutions $\Phi_{01}^-,\ \Psi_{01}^-$.
In order to investigate the behavior near $t=t_{1}$ and $t=t_{2}$ in a unified way, we introduce $\widetilde{\Phi}_{01}(z)$ and $\widetilde{\Psi}_{01}(z)$ by
\begin{eqnarray}
\pm i\widetilde{\Phi}_{01}&\equiv&\Phi_{01}^-
\ , \qquad
i\widetilde{\Psi}_{01}\equiv \Psi_{01}^- \ ,
\end{eqnarray}
The Borel transforms $\widetilde{V}_{1},\widetilde{U}_{1}$
of $\widetilde{\Phi}_{01},\widetilde{\Psi}_{01}$ satisfy 
\begin{eqnarray}
(e^{-p}-1)\widetilde{V}_{1}&=&\frac{k-1}{2}
(g'(p)+\frac{1}{2})+\frac{k-1}{4}(g(p)-1)+\widetilde{U}_{1}*g(p)
\nonumber
\\
(1-e^{p})\widetilde{U}_{1}&=&\frac{k-1}{2}f'(p)+\widetilde{V}_{1}*f(p)
\label{inn}
\end{eqnarray}
where
$$
B\left[\frac{e^{i\Phi_{00}}}{z}\right]=f(p)
,\ B\left[\frac{e^{\mp i\Psi_{00}}}{z}\right]=g(p)
$$

The power series expansions of $V_1,\ U_1$ is defined by
\begin{eqnarray}
\widetilde{V}_{1}=-\frac{\pm kt_{c}+1}{24}+\sum_{n=1}^{\infty}\widetilde{c}_{n} p^{n}
\ , \qquad 
\widetilde{U}_{1}=\frac{k(\pm t_{c}+1)}{24}+\sum_{n=1}^{\infty}\widetilde{d}_{n} p^{n}
\label{exp}
\end{eqnarray}
(\ref{inn}) and (\ref{exp}) gives the following form:

\begin{eqnarray}
\widetilde{c}_{n}=\pm kt_{c}\widetilde{c}^{(1)}_{n}+k\widetilde{c}^{(2)}_{n}+\widetilde{c}^{(3)}_{n}	
\ , \qquad
\widetilde{d}_{n}=\pm kt_{c}\widetilde{d}^{(1)}_{n}+k\widetilde{d}^{(2)}_{n}+\widetilde{d}^{(3)}_{n}
\label{ex}
\end{eqnarray}
where $\widetilde{c}^{(i)}_{n},\ \widetilde{d}^{(i)}_{n}$ are 
independent of $k$.
By substituting (\ref{ex}) into (\ref{inn}), we numerically get the following 
estimation.
$\widetilde{c}_{n},\ \widetilde{d}_{n}$ as follows:
\begin{eqnarray}
\widetilde{c}_{2n}&=&
B_{2}(k-1)\frac{(2n+2)(2n+1)}{(2\pi)^{2n}(-1)^{n}}
-(\pm B_{1}kt_{c}-B_{3}(k+1))\frac{2n+1}{(2\pi)^{2n}(-1)^{n}}
\nonumber
\\
\widetilde{d}_{2n}&=&
B_{2}(1-k)\frac{(2n+2)(2n+1)}{(2\pi)^{2n}(-1)^{n}}
-(\pm B_{1}kt_{c}-B_{3}(k+1))\frac{2n+1}{(2\pi)^{2n}(-1)^{n}}
\nonumber
\\
\widetilde{d}_{2n+1}&=&
-B_{4}(k-1)\frac{(2n+2)}{(2\pi)^{2n+1}(-1)^{n}}
\nonumber
\end{eqnarray}
This estimation gives (\ref{sol:numerical1}).

\end{document}